\documentclass{article} 
\usepackage{latexsym} 
\usepackage{graphicx} 
\usepackage{epsfig} 
\newcommand \be {\begin{equation}} 
\newcommand \bea {\begin{eqnarray}} 
\newcommand \ee {\end{equation}} 
\newcommand \eea {\end{eqnarray}}

\begin{document} 
\topskip 2cm 
\begin{titlepage} 
\rightline{\today} 

 \begin{center} 
{\Large \bf Gluon condensation and deconfinement critical
density  in nuclear matter}\\ 
\vspace{2.cm} 
{\large  Marcello Baldo$^{1,2}$,  Paolo Castorina$^{2,1}$, Dario Zappal\`a$^{1,2}$} \\ 
\vspace{.5cm}
{\sl $^{1}$ INFN, Sezione di Catania,  Via S. Sofia 64, I-95123, Catania, Italy} \\ 
\vspace{.2cm} 
{\sl $^{2}$ Dipartimento di Fisica, Universit\`a di Catania} \\ 
{\sl Via S. Sofia 64, I-95123, Catania, Italy}\\

\vspace{1.5 cm} 
{\sl e-mail: marcello.baldo@ct.infn.it; ~paolo.castorina@ct.infn.it; ~dario.zappala@ct.infn.it}\\ 
\vspace{2cm}

\begin{abstract} 
An upper limit to the critical density for the transition to the deconfined
phase, at zero temperature, has been evaluated by
analyzing the behavior of the gluon condensate in nuclear matter.
Due to the non linear baryon density effects, the  upper limit to the
critical density,$\rho_c$
turns out about nine times the saturation density,  $\rho_0$ for the value of the 
gluon condensate in vacuum  $<(\alpha_s/\pi) G^2>=0.012$ $GeV^4$.  For neutron
matter $\rho_c \simeq 8.5 \rho_0$. The dependence of the critical density
on the value of the gluon condensate in vacuum is studied.
\end{abstract} 
\end{center}

\vspace{2 cm}


\end{titlepage} 
 
\newpage

\section {Introduction}

The future experimental results both in relativistic heavy ions collisions and in astrophysics will give
a definite answer on the observation of new states of matter under extreme conditions.
Our understanding of particles and nuclear phenomena suggests that at finite temperature there is a phase transition
to deconfined gluons and massless quarks. QCD lattice calculations \cite{uno} strongly support the conclusion that
at a  critical temperature, $T_c$, a transition to the quark-gluon plasma phase occurs  and $T_c$ is equal to $T_ \chi$,
the critical temperature for the restoration of the chiral symmetry.
More recently it has been proposed that, at  high density and low  temperature,  QCD exhibits
a new phase due to a diquark condensation, called color superconductivity, with interesting astrophysical effects \cite{due}.
The understanding of the whole phase diagram of the theory requires an analysis of the dependence of the
order parameters on  the temperature and on the density.  Unfortunately   lattice calculations at large density are 
problematical although recently two  methods  have been proposed to overcome, at least for not too large values of the 
chemical potential, the technical difficulties that arise in lattice simulations \cite{tre}.

The regime, of low  $T$ and high  density, is particularly interesting not only from the theoretical point
of view but also because it has strong implications in astrophysical systems such as the neutron stars.
For example, the mass and the relation between the mass and the radius of a neutron star depends
on the state of quark matter in the core ( if any) and on the transition from nuclear matter to quark matter \cite{quattro}.

The value of the critical density at which this possible transition occurs inside the neutron star has been determined, 
up to now, by comparing and matching the two Equations of State (EoS), one for the nucleon matter and one for the 
deconfined quark matter, which are derived within two different frameworks. 
In this procedure a first order phase transition is assumed, and the Gibbs construction, or a modification  of it, is employed.

For the nucleon EoS, either microscopic or phenomenological approaches have been followed. 
In the microscopic approach, the nucleonic EoS is derived from a realistic nucleon-nucleon (NN) interaction, which 
fits the NN phase shifts and deuteron data,  by solving accurately the many-body problem for infinite nuclear matter. 
The main requirement is the  reproduction of the empirical saturation point of symmetric nuclear matter, as extracted 
from the Weiz\"acker-like mass formula for finite nuclei, and the compressibility at saturation as extracted 
from the analysis of monopole vibrations in nuclei. 
As a rule, three-body forces have to supplement the two-body forces in order to fulfill these requirements. 
In the purely phenomenological approach, the same requirements are used to fix or to 
constrain the parameters of the model \cite{cinque}.

For the quark phase, only simple models can be used, since no
``{\it ab initio}"  QCD calculations at high baryon density is
available. Both MIT bag model and the NJL model have been used.
Due to the uncertainties inherent to the models, the precise location
of the phase transition is not well known.

Moreover another constraint on the value of the critical density, is the phenomenological
observation that in heavy ion collisions at intermediate energy
( $ 10  MeV/A < E/A < 200  MeV/A $ ) no evidence of 
a transition to the quark-gluon plasma phase has been found. 
Indeed, all microscopic simulations 
that are able to reproduce a great variety of experimental data, do not need the introduction of such a transition. 
In these simulations the calculated nucleon density can reach values which are
at least 2-3 times larger than the saturation density $\rho_0$.
One can, therefore, conclude that symmetric or nearly symmetric nuclear  matter at few $MeV$ of 
temperature  does not exhibit any phase transition to the deconfined matter up to this baryon density. 
It has to be noticed that the phase transition in symmetric matter can occur at a
different baryon density than in neutron star matter,
where nuclear matter is closer to neutron matter than to symmetric
nuclear matter.

This constraint coming from heavy-ion physics appears
as an independent one, that should be fulfilled by any theory or model
of deconfinement. Indeed, the quark matter model mentioned above
seem to have, in some cases, serious difficulties to fulfill the
constraint (the transition occurs at too low density, or it does not occur 
at all), although it  produces `` reasonable " results for neutron
stars, where the transition does occur.
According to the previous discussion, it would be, therefore, desirable
to formulate an independent and differently founded estimate of the transition
point (or the possible mixed region).

In this Letter we shall address this problem  by discussing the dependence of 
the QCD gluon condensate \cite{sei}  
\be
\label{eq:1}
< {\alpha_s \over \pi} G_{\mu \nu}^a (0) G^{\mu \nu a} (0)> = M^4 \neq 0
\ee
on the density $\rho $   (at zero temperature),  as a signal of confinement-deconfinement transition.
In fact  the trace anomaly decreases with temperature \cite{aggi1} but does not vanish above the critical temperature
and then the gluon condensate is not, by itself, an order parameter for the transition.
However, the vanishing of the gluon condensate at large density gives an upper limit to the critical density
for the following reasons:

1) According to Dosch  \cite{sette} the string tension $\sigma$  is given by
\be
\label{eq:2}
\sigma \simeq M^4 \, \int d \tau dr f( {{- \tau ^2 - r^2} \over a^2})
\ee
where $a$ is the correlation length for the gluon field defined by
\be
\label{eq:3}
< {\alpha_s \over \pi} ~G_{\mu \nu}^a (x) G^{\mu \nu a} (0)> = M^4 f(x^2/a^2)
\ee
By Equation (\ref{eq:2}), it turns out that the string tension in proportional to
\be
\label{eq:3bis}
\sigma \simeq M^4  a^2
\ee
This implies that the behaviour of the string tension at finite density depends
not only on the depletion of the gluon condensate, $M^4$,
but also on the decreasing of the gluon correlation length at large density.
Therefore the vanishing of the gluon condensate
can give only an upper limit to  the critical density.

2) The gluon condensate is the sum of two 
independent terms, $ <E^2>$  and $<B^2>$, see for instance the review in \cite{digi}.
At finite temperature, $<E^2>$  vanishes
at the deconfinement transition, while $<B^2>$
remains finite even above the critical point and one can 
expect an analogous behavior at finite density \cite{adami,koch,brown}.
Again, this is an indication that the full gluon condensate  
provides an upper limit to the critical density.

3) As discussed by Hatta and Fukushima \cite{hatta}, the vanishing of the
 glueball can describe the critical phenomena at the deconfinement transition. 
But the glueball mass is usually considered proportional to the trace anomaly,
\be
\label{eq:3tris}
M_G \simeq < {\alpha_s \over \pi} ~G_{\mu \nu}^a (0) G^{\mu \nu a} (0)>^{1/4} =
M
\ee
which persists at any temperature. Then there is an apparent contraddiction solved 
by considering that only a fraction of the glueball
 mass comes from the trace anomaly ( see \cite{rothe} ) 
and that, as previously commented, only the "electric" glueballs
 (those which contains time-like links) become massless at the phase transition
while the "magnetic" glueballs remain massive. Then the vanishing of the glueball
mass occurs at a smaller density than the estimate based on the gluon condensate.

According to the above issues, we shall evaluate the upper limit to the critical density 
by analyzing the dependence of the gluon condensate on the baryon density.
In Section 2 we shall give an heuristic argument for the occurence of the transition.
in Section  3  the known results about the behavior of the gluon condensate in
nuclear matter are  recalled and we discuss a phenomenological consequence of this behavior;
Section 4  is devoted to the calculation of the critical density including the non linear effects 
and Section 5 to the conclusions.

\section  {TOY CALCULATION}

The glueball mass is related to the gluon condensate. Then the first heuristic argument 
about the critical density can be obtained by looking at the behavior of the glueball mass as a function of the
baryon density. In this Section we shall consider a  simple calculation to give some very preliminary indications
of the deconfinement transition at large density.

Let us consider the color potential given by a linear rising term plus the perturbative contribution
\be
\label{eq:4}
V(r)= \sigma r - {4 \over 3}{\alpha \over r}
\ee
where $\sigma$ is the effective string tension, $\alpha$ is an effective constant (independent on the scale) and 
the factor $4/3$ is due to the color degrees of freedom.
Let us assume that the glueball of total energy $E$ is a bound state of two massless gluons:
\be
\label{eq:5}
E= 2 p + \sigma r - {4 \over 3}{\alpha \over r}
\ee
From the uncertainty principle $p r \geq 1$ and then,
by minimizing $E$  with respect to $r$, it turns out that $\sigma r_0^2 = (2-4 \alpha /3) $ and $E(r=r_0)= E_0 = 2 \sigma r_0$. For (almost)
realistic values $\sigma \simeq 0.3~GeV^2$ and $\alpha \simeq 0.1$, $E_0 \simeq 1.5~GeV$ not far from the present glueball mass limit.

This simple picture is modified by  baryon density effects ( which will be  discussed later)
and in the medium the potential in Eq. (\ref{eq:4}) is replaced by  the screened potential \cite{otto}
\be
\label{eq:7}
V_s(r)= \sigma r ~{{ 1 - {\rm e}^{-\mu r}} \over {\mu r}} - {4 \over 3}{\alpha \over r}~ {\rm e}^{-\mu r}
\ee
where $\mu$ is the screening parameter and  the screened potential has the asymptotic limit $V_s(\infty)=\sigma / \mu$.
By repeating the previous calculations, it turns out that $r_0=r_0(\mu)$ , 
and the value of the minimum energy is
\be
\label{eq:9}
E_0 - {\sigma \over \mu} =  {\rm e}^{-\mu r_0} \left \lbrack \sigma \left ( r_0 - {1 \over \mu}\right ) + {4 \over 3}\alpha \mu \right \rbrack
\ee
which  shows that there is a critical value of the screening parameter $\mu _c$ at  which the bound state disappears.

To evaluate the dependence of $E_0$ on the baryon density $\rho$ (and then the critical density)
we need to know its relation with the screening parameter.
By assuming a Debye screening due to a 2 flavor massless quarks gas of density $\rho_q = 3 \rho$ \cite{nove}, then
\be
 \label{eq:10}
\mu ^2 =  {12 \alpha \over \pi}   \left ( 3 {\pi ^2 \over 2}  \rho\right )^{ 2 \over 3}
\ee
By means of  Eqs. (\ref{eq:7}) and  (\ref{eq:9}), it turns out that the critical value  where the bound 
state disappears is  $\mu_c \simeq 0.24~ GeV$ that corresponds to a baryon density  $\rho  \simeq 3 \rho_0$
where $ \rho_0 \simeq 0.16 ~ {\rm fm}^{-3}$ is the saturation density.

The previous argument represents only a qualitative way to show that the glueball disappears from the spectrum
at some critical density and does not pretend to make any quantitative statement. For example,
if one fixes the value of the parameter $\sigma$ in the potential $V(r)$ to fit the glueball mass
$m_g \simeq 1.7~GeV$, then it turns out that the critical density is about five times the saturation density. 

\section{GLUON CONDENSATE IN NUCLEAR MATTER}

In this Section we recall the main results concerning the behavior of the gluon condensate in nuclear matter (see \cite{undici}).
In vacuum, at $T = 0$, the estimates of the gluon condensate \cite{dodici} indicate the range
\be
\label{eq:11}
0.003 ~ GeV^4 < \left ( <  {\alpha_s \over \pi} G_{\mu \nu}^a (0) G^{\mu \nu a} (0)> \right )  < 0.021 ~GeV^4
\ee
By defining the gluon condensate at zero temperature and  at finite density in nuclear matter
\be
\label{eq:12}
g(\rho) = ~< M |{\alpha_s \over \pi} G_{\mu \nu}^a (0) G^{\mu \nu a} (0)| M >,
\ee
the non linear density effects are taken into account by considering  \cite{undici}
\be
\label{eq:14}
g(\rho) = g(0) -{8  \rho \over 9} \left (m + \epsilon (\rho)- \sum m_i <N| \overline  q_i q_i|N>\right )
\ee
where  $m$ is the nucleon mass,
$m_i$ are the current quark masses, $<N| \overline  q_i q_i|N>$ is the fermionic condensate
in the nucleon in the medium, the gluon condensate in vacuum is indicated with $g(0)$, 
$\epsilon (\rho)$ is the binding energy per nucleon in the medium and 
the sum of the fermionic condensates is extended only to the three light flavors 
\cite{undici,undiciagg}.

According to previous equation the gluon condensate at finite density decreases by increasing $\rho$ and,
despite  this result refers to nuclear matter, one can ask if  some phenomenological indication of this depletion
comes out,  for instance, by comparing light and heavy nuclei.
To answer to this question one can consider the behavior of the deep inelastic scattering structure function,
$F_2^A$, per nucleon in a nucleus.

For very small values of the scaling variables $x$ the experimental data on the  ratio of the structure functions
for two different nuclei,  $F_2^A/F_2^B$, shows
a depletion (shadowing).

In the small $x$ region the structure function is described by the pomeron exchange \cite{tredici}
and, indeed, for $x \rightarrow 0$ the structure functions are related
to the quark-pomeron effective coupling in the target (T), $\beta_T$, 
\be
\label{eq:15}
F_2(x) _ {\longrightarrow\atop {x \rightarrow 0}} \; \beta_T
\ee
and the ratio of the structure functions per nucleon in the nucleus A with respect to nucleus B is
\cite{quattordici} 
\be
\label{eq:16}
{{{F_2(x)^A \over {F_2(x)^B}}} \vert _  {\longrightarrow\atop {x \rightarrow 0}} \; {\beta_A \over {\beta_B}}}
\ee

On the other hand, in the model proposed by Landshoff and Nachtmann \cite{quindici}, the pomeron exchange is described by 
nonperturbative gluons and  the quark pomeron coupling is related to the gluon condensate according to relation \cite{sedici}
\cite{diciasette}
\be
\label{eq:17}
\beta_T \simeq a^5 < T |{\alpha_s \over \pi} G_{\mu \nu}^a (0) G^{\mu \nu a} (0)| T >
\ee

In this analysis one can consider the correlation length $a$ as independent of the target, because
it is related to the pomeron coupling to off-shell quarks \cite{sedici}
\cite{diciasette} and therefore one obtains
\be
\label{eq:18}
{{{F_2(x)^A \over {F_2(x)^B}}} \vert _  {\longrightarrow\atop {x \rightarrow 0}} \; {\beta_A \over {\beta_B}}}
= { { < A |{\alpha_s \over \pi} G_{\mu \nu}^a (0) G^{\mu \nu a} (0)| A >} \over
{< B |{\alpha_s \over \pi} G_{\mu \nu}^a (0) G^{\mu \nu a} (0)| B >}}
\ee
Then the  last ratio can be written as
\be
\label{eq:19}
{{F_2(x)^A \over {F_2(x)^B}}} \simeq {g(\rho _A)\over g(\rho _B)}
\ee
where $\rho _A$ and $\rho _B$ are the nuclear densities of $A$ and $B$.
For example the experimental data on the structure function at small $x$
give a ratio $F_2(x)^{Fe} / F_2(x)^C$ of about $0.9$ at $x=0.007$ \cite{diciotto}.

One can conclude that, according to the model in \cite{quindici},   there is at least an indication that
at zero temperature  the gluon condensate decreases in going from light to heavy nuclei.
Of course the calculation of the numerical value
of the shadowing suppression is beyond the scope of this paper.
Indeed we are considering  the nuclear matter case and not finite nuclei and moreover a quantitative prediction of
the low $x$ behavior of the structure functions, even in the case of  the free nucleon case, needs a deep 
understanding of the transition between the perturbative and the nonperturbative regimes in QCD.

\section {NON LINEAR EFFECTS}

The indications provided in the previous Section suggest that  Eq. (\ref{eq:14})  could be a good starting point to 
evaluate the critical density by introducing the non linear effects in $\epsilon$ ($\rho$). This requires the
analysis of the dynamical description of nuclear matter.

At low baryon density and zero temperature nuclear matter can be
described as a dilute gas of nucleons. The residual
nucleon-nucleon interaction is mainly mediated by mesons. In this regime
it is a reliable approximation to model nuclear matter as a gas
of nucleons interacting through a static potential. The latter
can be  extracted from the phenomenological analysis
of nucleon-nucleon scattering data. Once the static nucleon-nucelon
interaction is given, the many-body problem can be accurately
solved  and the energy density as a function of baryon
density can be calculated ( see eg. \cite{mar1}). 

Only at very low density, much lower than
the saturation density, the energy density follows a linear behavior, 
since different correlations start to develop in the system
as the density increases. Therefore, the mean field approximation turns
out to be inadequate. In particular, Pauli principle strongly modifies
the two-nucleon scattering process in the medium and this effect
is density dependent. Other many-body effects, like the momentum
dependence of the single particle potential and three-body correlations
contribute to the nonlinear terms. Much work has been done
along these lines, and the EoS of nuclear
matter,  up at least to saturation density,  can be considered
well established \cite{mar1}. 

Near saturation density a non-negligible contribution
of three-body forces and/or relativistic effects appear necessary
to get an accurate saturation point.  Their contributions is anyhow
at few percents levels and do not affect the main trend of the
energy density and of the corresponding nonlinear terms.

At higher density the calculations contain a certain degree of
extrapolation, since both two-nucleon and three-nucleon forces
must be extended beyond the values of the relative momenta where
they have been phenomenologically checked. However, up to density
2 - 3 times larger than the saturation density the nuclear matter
EoS obtained along these lines can be considered
still reliable.

The calculation of $\epsilon$ ($\rho$) following the previous indications 
has been performed in \cite{mar2}. The results are displayed  in Table 1
of  \cite{mar2} for two different potentials which include three-body forces
both for symmetric and neutron matter and we do not report them here.
From these values and from Eq. (\ref{eq:14}), one obtains an 
approximated dependence of the gluon condensate on the baryon density.
The quark condensate contribution, which as already noticed
is restricted to the light flavors, is small because of the  size of the quark masses 
with respect  to the nucleon mass which also appears in  Eq. (\ref{eq:14}).
Our calculation is performed in the chiral limit  with $m_u=m_d=0$ and we have included the strange 
contribution taking  $m_s=0.1~GeV$ and $<N| \bar s s |N> =1$, according to the estimates
quoted in \cite{undici} and the overall effect of the strange condensate is only few percent
on the estimate of $g(\rho)$. In Eq. (\ref{eq:14}) a possible non-linear contribution due to
the strange quark condensate has been neglected because the included linear term 
is small and one expects that this further correction should be even smaller since
the strange meson exchange plays a negligible role in N-N interaction, 
as discussed in \cite{undici}.

According to the magnitude of the condensate in vacuum ( see Eq. (\ref{eq:11}))  
one finds different values of the upper limit to the critical density. 
The behavior of the condensate  as a function of the  density 
is depicted in Figure 1 where  $g(\rho)$ is plotted for three representative 
values of  $g(0)$ and with   $\epsilon$ ($\rho$)
determined in \cite{mar2} in the case of symmetric matter for  the two  potentials AV14 and Paris.
Figure 1 remarkably shows the importance of the nonlinear effects. Indeed the curves that 
would be  obtained by neglecting $\epsilon(\rho)$,  correspond
approximately  to the straight lines tangent to the curves in Fig. 1 at the lowest value
of $\rho$. These curves would yield a much larger critical density (above $2.5~ fm^{-3}$).
The nonlinear effects due to $\epsilon(\rho)$ reduce the critical density  of almost  a factor two.
In the case of neutron matter the behavior is analogous 
to the one shown in Fig. 1 and,  as expected, a smaller  critical density is found.

In Figure 2  the critical density is  plotted as a function of the magnitude 
of the gluon condensate in vacuum both for symmetric and neutron matter.
In each case we have displayed two curves corresponding again to the two potentials
considered in  \cite{mar2}.
Our upper limits are compatible with the values of the critical density  suggested in \cite{cinque}.

\section{CONCLUSIONS}

In summary, we have considered a way of estimating the upper limit to the critical density  
for the transition to the deconfined
phase at zero temperature by taking into account, in the determination of the gluon condensate  in nuclear matter,
the effects of the binding energy per nucleon which have already been  determined  by means of the EoS for nuclear matter.
Points 1) - 3) in the Introduction clearly explain that the gluon condensate is not an order parameter and 
that the present approach can only give an upper limit to the critical density. On the other hand, our
analysis shows that there are good indications that  Eq. (\ref{eq:14}), once the nonlinear effects have been included,
can reasonably describe the gluon condensate dependence  on the baryon density at zero temperature.

The  quantitative analysis  at large baryon density gives  for the (central)  value of the gluon condensate in vacuum  
$g(0)=0.012$ $GeV^4$, an upper limit  $\rho_c \simeq 9.1 \rho_0$.
Being an upper limit, this value is  compatible 
with the typical values obtained by using phenomenological models for the quark matter
and the Gibbs construction \cite{quattro,cinque}. A  lower
value of the vacuum gluon condensate, $g(0) \approx 0.008-0.01$, which
is well within the theoretical uncertainty, would produce a critical 
density closer to these phenomenological estimates. 
Despite these uncertainties, the result appears 
highly non-trivial, since the method employed here is totally different. 
The decreasing of the critical density for asymmetric matter  ( $\rho_c \simeq 8.5 \rho_0$),
is less than expected on the basis of detailed dynamical calculations
\cite{muller} but it is remarkable that this effect is nevertheless obtained in a totally different  framework.

\vspace{0.6 cm} 
\leftline{\bf Acknowledgements} 
The authors thank A. Drago for useful discussions. 
P.C. and D.Z. acknowledge
the MIT Center for Theoretical Physics for kind hospitality.
P.C. has been partially supported by the INFN Bruno Rossi exchange program.

\vfill
\eject

\begin{figure}
\epsfig{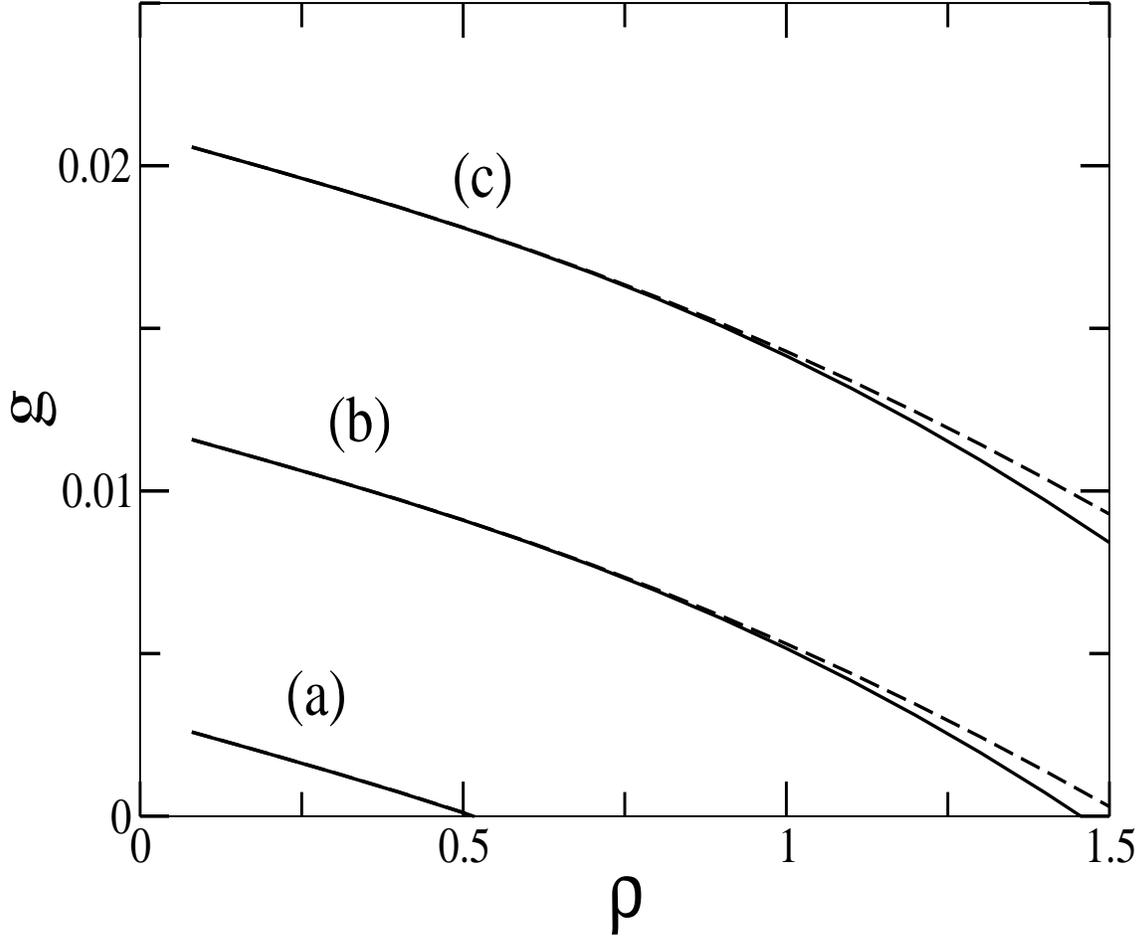}
\caption{$g(\rho)$ in $GeV^4$ {\it vs.} the density $\rho$ in ${\rm fm}^{-3}$
for  $g(0)=.003~ GeV^4$ (a),  $g(0)=.012~ GeV^4$ (b),  $g(0)=.021~ GeV^4$ (c).
Dashed and solid lines correspond to the two different determinations  of $\epsilon(\rho)$
in \cite{mar2}, obtained respectively with the AV14 potential and the Paris potential,
for symmetric matter (see text).
}
\end{figure}

\vfill
\eject

\begin{figure}
\epsfig{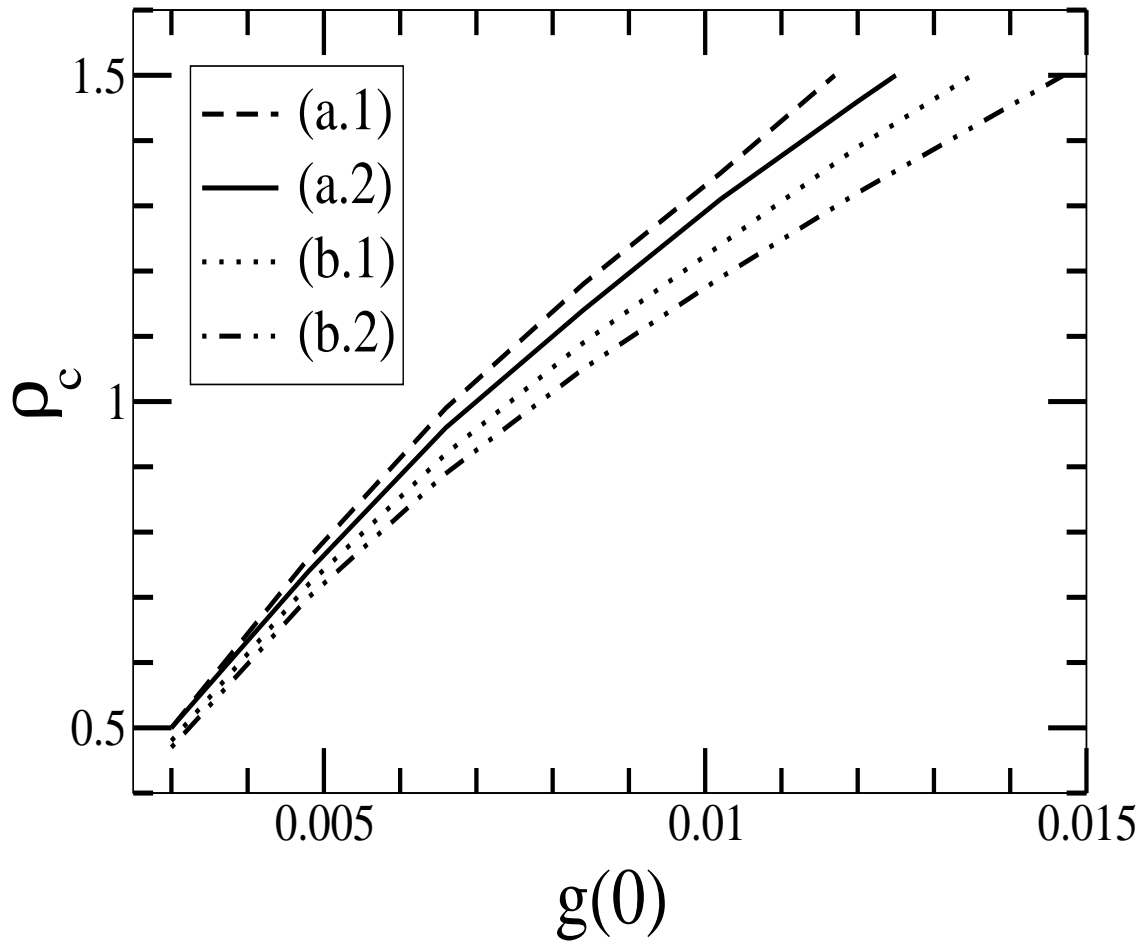}
\caption{ The critical density $\rho_c$ in ${\rm fm}^{-3}$ as a function of 
the condensate in vacuum $g(0)$ in $GeV^4$.  Curves a.1 and a.2 correspond 
to the two  determinations  of $\epsilon(\rho)$, 
obtained respectively with the AV14 potential and the Paris potential,
 for symmetric matter as in Figure 1,
and  b.1 and b.2 the same for neutron matter.
}
\end{figure}
\end{document}